# Theory of CHTW-systems. Born by Petri Nets

**Alexander Yu. Chunikhin**

Palladin Institute of Biochemistry

National Academy of Sciences of Ukraine

alexchunikhin61@gmail.com

https://orcid.org//0000-0001-8935-0338

**Abstract**.

For the first time, the concept of CHTW-systems as a multidimensional representation of Petri nets, based on the assumption of the spatial distribution of tokens (resources) in positions (branes) and, accordingly, the spatial representation of transitions and arcs is proposed. The theoretical constructs are based on the concept of hybrid functional Petri nets [10]. The introduced concepts of *branes* and *carriers* are distant analogies of the corresponding concepts in superstring theory, but the theory of CHTW-systems is neither part of superstring theory nor its development. The description of CHTW-system as a dynamic system for stationary and non-stationary cases is considered. The initial classification of CHTW systems is provided enabling understanding of further research directions.

**Keywords**: Petri nets, C-brane, T-brane, H-carrier, W-carrier, mark-function, CHTW-system.

1. **Introduction**

Petri nets, first introduced by K.A. Petri in 1962 [13], are a well-known apparatus for modeling of discrete systems. Currently, there are a significant number of extensions to the basic concept of Petri nets. In addition to discrete ones, continuous, hybrid (discrete-continuous) [6], and hybrid functional Petri nets [10] have been developed. Timed [14] and Coloured Petri nets [7], as well as nets that take into account one or another aspect of uncertainty namely stochastic [3], fuzzy [1, 8], and rough [11], are widely used. Structural variations are reflected in the concepts of nested [9], open [2], hierarchical and reconfigurable Petri nets [12], as well as in the concept of Petri nets' receptors and effectors [5]. The fundamentals of creative Petri nets theory [4] provide an understanding of the possibility of fusion and defusion processes modeling, generation and destruction of substructures within an integral Petri net structure. Despite the obvious diversity, all these types of Petri nets can be called *networks with point parameters*. Petri nets marking is also essentially attributing a certain cardinal number to each network position.

At the same time, most real systems are complex dynamic spatially distributed systems, often non-stationary. In such systems, both resources and processes have the form of a functional distribution on some (often more than one) multidimensional space.

Is it possible, within the framework of the formalism of Petri nets, to represent the distribution of tokens in a position as a function of many variables? How will the arcs and transitions in such nets change in this case?

It is obvious that changes in such "multidimensional" Petri nets must be introduced to both the names of net elements and their designations, as well as the conditions for "firing" of transitions and analytical expressions for the functioning of such nets. For example, a position "stretched" along two dimensions can no longer be called by the local, "point" name. This is no longer a position, but rather a local two-dimensional membrane. The same applies to multidimensional transitions and multidimensional arcs. That is why the basic concepts of a *brane* for multidimensional positions and transitions, and a *carrier* for multidimensional arcs were borrowed from the physical theory of superstrings [15]. The introduced concepts are distant analogies of the corresponding concepts in superstring theory. The theory of CHTW-systems is neither part of superstring theory nor its development.

## 2. Preliminaries

Informally, extension from traditional Petri nets to their multidimensional representation can be considered from two points of view.

A) Let's give to some set of *m* tokens in P position a *m(x)* spatial distribution along a given X axis, in a certain interval [a, b]∈X (Fig.1).

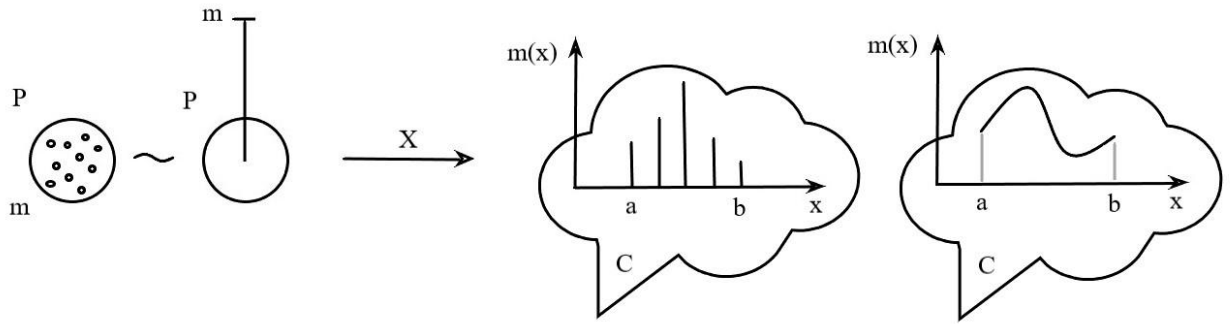

Fig.1

The P position seemingly stretches along the X axis and becomes a one-dimensional brane (C-brane), discrete or continuous. We can say that the initial P position is a 0-dimensional brane.

B) Imagine that a certain Petri net consists of a set of *n* homogeneous positions containing tokens (resources) of one same type. A given set of positions can be defined as a discrete one-dimensional C-brane (Fig. 2), and when tending to infinity ($n \to \infty$) as a continuous one-dimensional C-brane. The corresponding transitions form a T-brane, and the arcs form carriers of two different types.

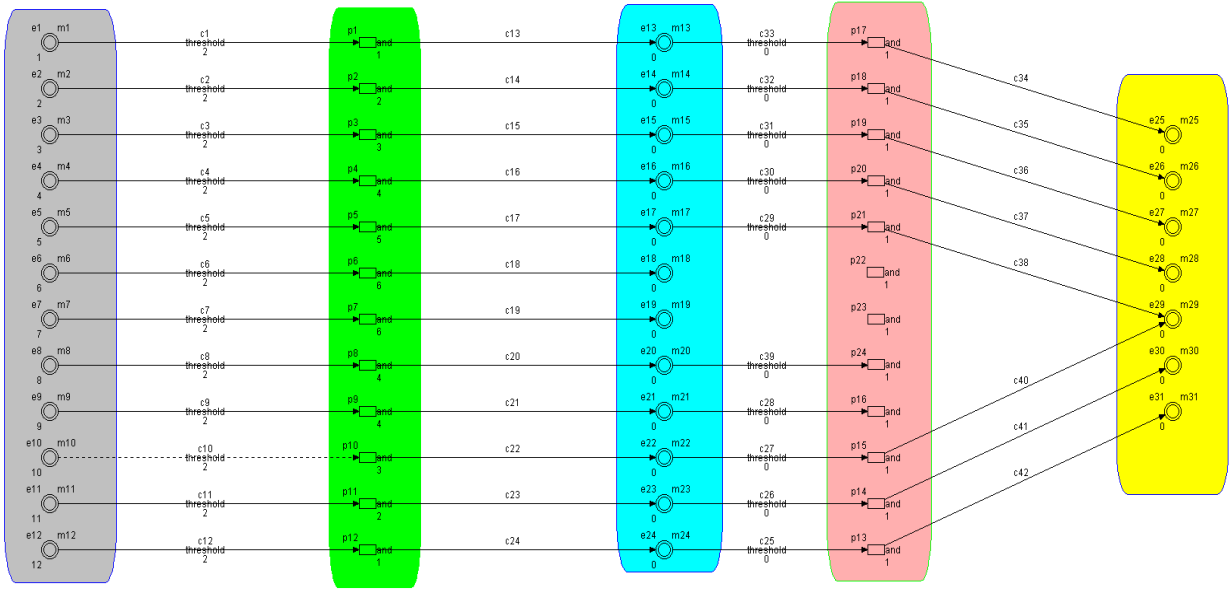

Fig.2

With natural extension to a larger number of dimensions, it is possible to form multidimensional C- and T-branes of arbitrary dimensions. Accordingly, the arcs connecting them are transformed into multidimensional carriers: *H* (threshold type) and *W* (transformative type). In accordance with these notations, we will call multidimensional extensions of Petri nets as *CHTW-systems*.

## 3. The concept of CHTW-systems

Let us introduce the basic concepts, definitions and notations of CHTW-systems, which allow to clearly distinguish them from traditional Petri nets.

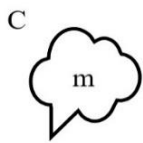

C-brane (a **c**ontainer brane) – a resource brane – accumulates, stores and releases distributed *m* resource.

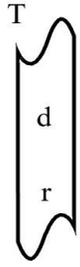

T-brane (a **t**ransition brane) — ensures the formation of "firing" (*d*-condition) for both taking out the resource from C-branes with *r* intensity and generation of a new resource.

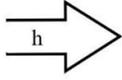

H-carrier (a **th**reshold carrier) – ensures the transfer of the resource from C-brane to T-brane in areas where *m* resource of the C-brane exceeds the value of the carrier *h* threshold.

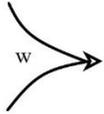

W-carrier (**w**hat/**w**here carrier) – a transformation carrier – ensures the formation of a new resource in C-brane associated with the corresponding T-brane.

M – integral (total) resource of CHTW-system – the totality of resources of all C-branes in the system.

***Definition 1***. We will call *CHTW-system* (Ξ) a fivetuple of the following form:

$$\Xi = (C, H, T, W, M),$$

where

$C = \{C_1, C_2, \ldots, C_p\}$ – a set of C-branes included in CHTW-system.

$T = \{T_1, T_2, \ldots, T_s\}$ – a set of T-branes included in CHTW-system.

$H = \{h_{1i}, h_{2j}, \ldots, h_{pq}\}$ – a set of threshold H-carriers from C-branes to T-branes in CHTW-system.

$W = \{w_{i1}, w_{j2}, \ldots, w_{sg}\}$ – a set of transformation carriers from T-branes to C-branes in CHTW-system.

$M = \{m_1, m_2, \ldots, m_p\}$ – a set of mark-functions that characterize resource distribution in C-branes in CHTW-system.

Since in each C-brane the mark-function of resource distribution on some *n*-dimensional space X: m(X), in general, changes with each step of CHTW-system operation, we will denote such a mark-function as m(X|k). The same reasoning is valid, in the general case, for the remaining components of CHTW-system. Then, for the simplest non-stationary CHTW-system (Fig. 3) we write down:

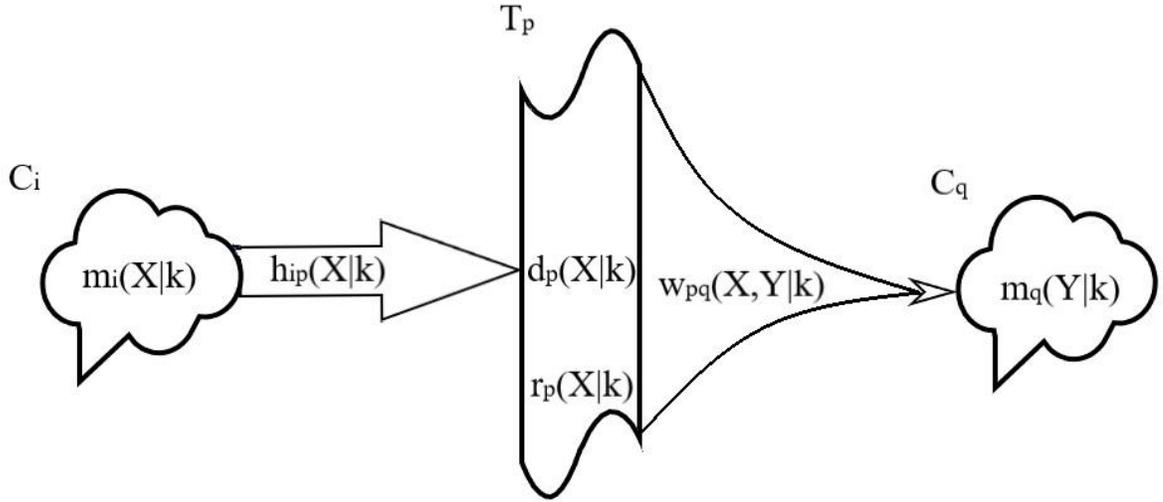

Fig.3

$m_i(X|k)$ – n-dimensional function of $m_i$ resource distribution on $X = (x_1, ..., x_n)$ space in the *i*-th C-brane at the *k*-th step of the CHTW-system operation;

$h_{ip}(X|k)$ – n-dimensional function of $h_{ip}$ threshold distribution on $X = (x_1, ..., x_n)$ space in H-carrier between the *i*-th C-brane and the *p*-th T-brane at the *k*-th step of the CHTW-system operation;

$r_p(X|k)$ – n-dimensional function of $r_p$ resource uptake intensity distribution from C-branes connected by H-carriers to the *p*-th T-brane on $X = (x_1, ..., x_n)$ space at the *k*-th step of the CHTW-system operation;

$d_p(X|k)$ – the "firing" function of the *p*-th T-brane – n-dimensional binary function of "effective" intervals distribution on $X = (x_1, ..., x_n)$ space at the *k*-th step of the CHTW-system operation;

$w_{pq}(X,Y|k)$ – the operator of transformation of the *n*-dimensional "firing" function of the *p*-th T-brane on $X = (x_1, ..., x_n)$ space into the *u*-dimensional function of $m_q$ resource distribution in the *q*-th C-brane on $Y = (y_1, ..., y_u)$ space at the *k*-th step of the CHTW-system operation:

$$w_{pq}(X, Y|k): d_p(X|k) \to m_q(Y|k).$$

Let us describe in more detail the procedure of determination of the $d_p(X|k)$ "firing" function of the $T_p$-brane. The "firing" of the $T_p$-brane is determined by the simultaneous fulfillment of two conditions. At each point $x \in X$, the $m_i(x)$ resource must exceed both the $h_{ip}(x)$ threshold of the corresponding H-carrier and the resource uptake intensity ($r_p(x)$) of the $T_p$-brane. This determines the following:

$\Delta_{ip}(X|k) = m_i(X|k) - h_{ip}(X|k)$ – the function of the resource excess in the $i$-th C-brane over the H-carrier threshold between the $i$-th C-brane and the $p$-th T-brane on $n$-dimensional $X = (x_1, ..., x_n)$ space at the $k$-th step of CHTW-system operation;

$\delta_{ip}(X|k) = m_i(X|k) - r_p(X|k)$ – the function of the resource excess in the $i$-th C-brane over the intensity of the resource uptake by the $p$-th T-brane on $n$-dimensional $X = (x_1, ..., x_n)$ space at the $k$-th step of the CHTW-system operation;

$\Theta(\Delta_{ip}(X|k))$ – the Heaviside function of $\Delta_{ip}(X|k)$ – a distribution of the fact that the resource in the $i$-th C-brane exceeds the H-carrier threshold between the $i$-th C-brane and the $p$-th T-brane on $n$-dimensional $X = (x_1, ..., x_n)$ space at the $k$-th step of the CHTW-system operation;

Here the Heaviside function is defined as $\Theta(x) = \begin{cases} 0, x \leq 0 \\ 1, x > 0 \end{cases}$;

$\Theta(\delta_{ip}(X|k))$ – the Heaviside function of $\delta_{ip}(X|k)$ – a distribution of the fact that the resource in the $i$-th C-brane exceeds the intensity of the resource uptake by the $p$-th T-brane on $n$-dimensional $X = (x_1, ..., x_n)$ space at the $k$-th step of the CHTW-system operation;

$d_{ip}(X|k) = \Theta(\Delta_{ip}(X|k)) \cdot \Theta(\delta_{ip}(X|k))$ – (partial) function of $T_p$-brane "firing" with $C_i$-brane resource.

For the case of multiple entry of two or more C-branes through H-carriers into $T_p$-brane, it can be shown that the integral "firing" function $(d_p(X|k))$ is the product of partial "firing" functions:

$$d_p(X|k) = \prod(d_{.p}(X|k)).$$

All of the above considerations allows us to formulate the following propositions.

***Proposition 1***. Branes of the same type are not directly connected by any carriers.

***Proposition 2***. Branes of different types can interact through carriers of the corresponding types.

***Proposition 3***. Any combination of brane bundles of *C-h→T* kind are necessarily realized on the same space for all components.

The proof follows from the procedure of "firing" function determination:

$d_{ip}(X|k) = \Theta(\Delta_{ip}(X|k)) \cdot \Theta(\delta_{ip}(X|k))$, and $d_p(X|k) = \prod(d_{.p}(X|k))$.

***Proposition 4***. Any number of W-carriers can emerge from one T-brane, generating any number of new C-branes on spaces of arbitrary dimension.

***Definition 2***. The *dimension of a brane* will be the dimension of the space on which the brane is defined.

From *Proposition 4* the below follows immediately**:**

***Proposition 5***. One CHTW-system can contain branes of different dimensions (Fig. 4).

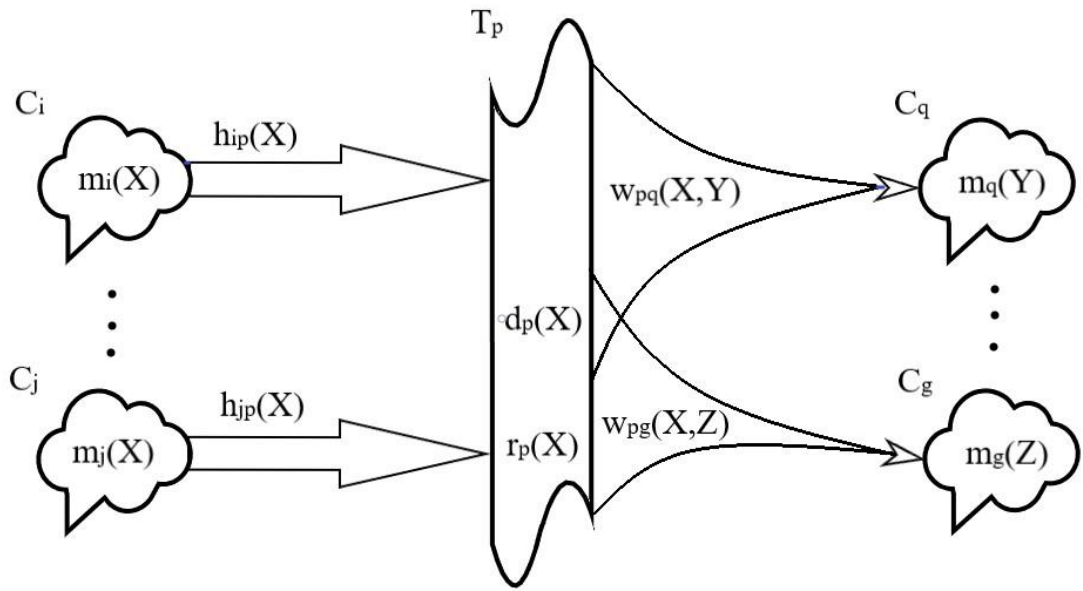

Fig.4

There are special types of H-carriers in CHTW-systems namely a *blocking* (inhibitory) and an *associative* (catalytic) carriers.

B-carrier (**B**locking carrier) – a carrier that blocks the "firing" of T-brane in areas where the $m$ resource exceeds $b$ threshold and does not use the resource of the corresponding C-brane.

A-carrier (**A**ssociative carrier) – a carrier that ensures "firing" of T-brane in areas where $m$ resource exceeds $a$ threshold, but does not use the resource of the corresponding C-brane.

A graphic representation of all types of carriers is shown in Fig. 5.

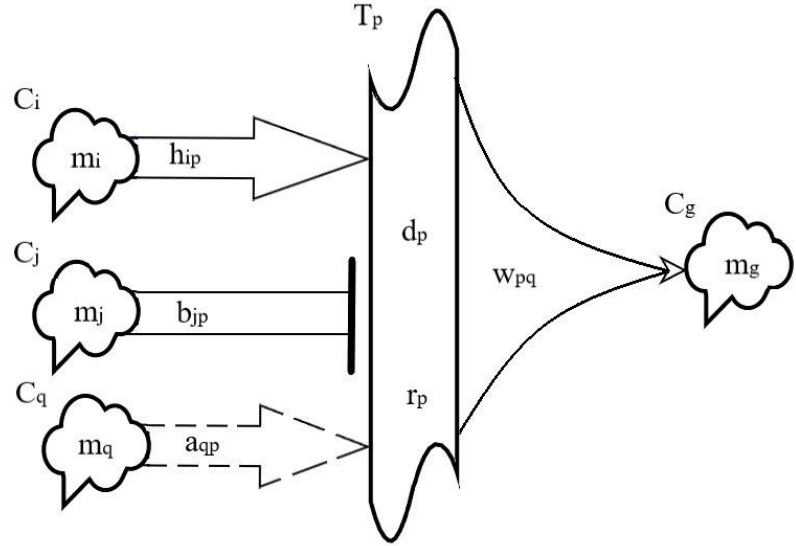

Fig.5

Let us define analytical expressions for the formation of T-brane "firing" for this functional scheme. Obviously, the condition for T-brane "firing" ($d_p$) will be three-component:

$$d_p = d_{ip} \cdot d_{jp} \cdot d_{qp},$$

where

$d_{ip} = \Theta(\Delta_{ip}) \cdot \Theta(\delta_{ip})$, где $\Delta_{ip} = m_i - h_{ip}$, $\delta_{ip} = m_i - r_p$;

$d_{jp} = \Theta(-\Delta_{jp})$, где $\Delta_{jp} = m_j - b_{jp}$;

$d_{qp} = \Theta(\Delta_{qp})$, где $\Delta_{qp} = m_q - b_{qp}$.

In this case, $m'$ mark-functions of C-branes after the activation of $T_p$-brane will take the following form:

$m_i' = m_i - r_p d_p$;

$m_g' = m_g + w_{pg} d_p$;

$m_j' = m_j$;

$m_q' = m_q$.

The above expressions exhaustively describe the behavior of CHTW-system at one operating step. For completeness of the description, it is desirable to obtain expressions that describe the general dynamics of the CHTW-system functioning, regardless of the specific characteristics of the branes.

## 4. CHTW-system as a Dynamical System

Let us first consider a stationary CHTW-system, the H and |r> parameters of which are constant during the system functioning. We introduce the following notation, invariant to the special characteristics of branes and carriers:

|m(k)> - resource vector-function [dim C × 1] in C-branes at the *k*-th step of the CHTW-system operation;

$\mathbf{H}$ = <h> - H-carriers matrix [dim C × dim T] of CHTW-system;

|r> - rate vector-function [dim T × 1] of CHTW-system;

|d(k)> - vector-function [dim T × 1] of T-branes "firing" at the *k*-th step of CHTW-system operation;

$\mathbf{W}$ = <w> - W-carrier matrix (operator) [dim T × dim C] of CHTW-system.

It is also advisable to define *connectivity matrices* of the following form:

$S_H$ = < $1_{ct}$ > - C-branes →T-branes connectivity matrix [dim C × dim T] of CHTW-system;

$S_W$ = < $1_{tc}$ > - T-branes → C-branes connectivity matrix [dim T × dim C] of CHTW-system.

Then the $\mathbf{R_S}$ = <$r_{ct}$> *resource uptake matrix* [dim C × dim T] is obtained by replacing the units of $S_H$ connectivity matrix with the values of the intensity of resource uptake by the corresponding T-brane.

$$S_H \rightarrow \mathbf{R_S}: 1_{ij} \rightarrow r_j, \forall i, j.$$

The dynamics of changes in mark-functions (resources) in CHTW-system is described by the following equation:

$$|m(k+1)> = |m(k)> - \mathbf{R_s} |d(k)> + \mathbf{W^T} |d(k)>.$$

For the case of a non-stationary CHTW-system: $\mathbf{H}(k)$ = <h(k)>, |r> = |r(k)>, $\mathbf{W}(k)$ = <w(k)>, and

$$|m(k+1)> = |m(k)> - \mathbf{R_s}(k) |d(k)> + \mathbf{W^T}(k) |d(k)>.$$

If there are inhibitory and associative carriers in the structure of CHTW-system, the following rule takes place.

***Rule.*** In $\mathbf{R_s}$ matrix, the rows corresponding to A- and B-carriers are zero rows.

This rule is a consequence of the functional purpose of these carriers – when T-brane is "fired" the available resource in C-branes corresponding to these carriers is not used, but can be replenished. When solving specific problems, the obtained expressions must be supplemented with appropriate spatial components taking into account their consistency (see *Propos.3*).

*Example*. Let a stationary CHTW-system with feedback containing branes of three different spaces (X, Y, Z) be described by the following functional diagram (Fig. 6).

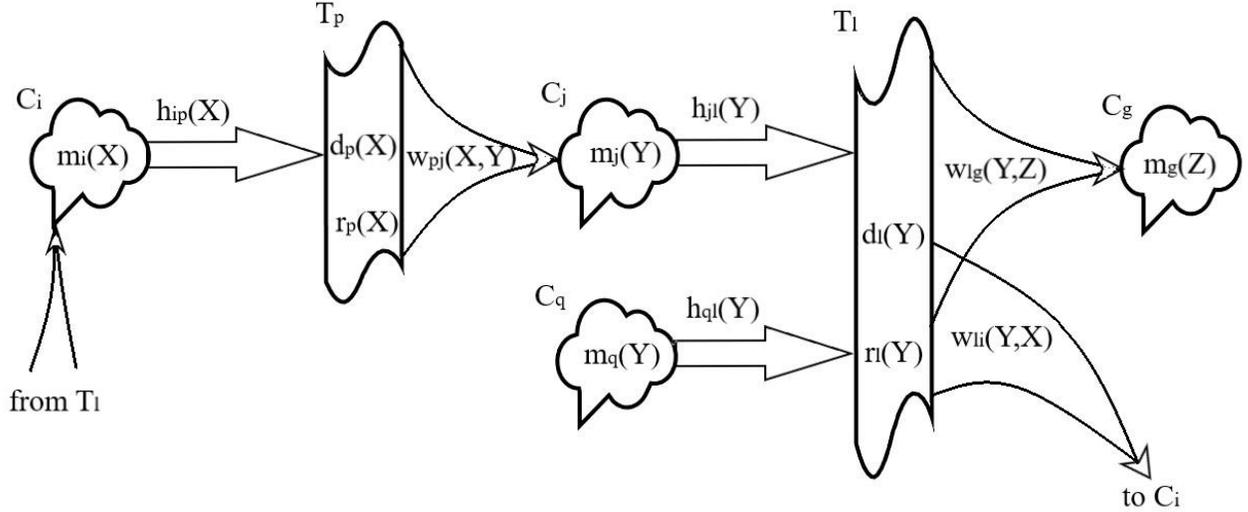

Fig.6

Then from

$$|m(k+1)> = |m(k)> - \mathbf{R_s}|d(k)> + \mathbf{W^T}|d(k)>$$

we can get the expression

$$|m(k+1)> = |m(k)> + (\mathbf{W^T} - \mathbf{R_s})\,|d(k)>,$$

where

$$|m(k)> = \begin{pmatrix} m_i(X|k) \\ m_j(Y|k) \\ m_q(Y|k) \\ m_g(Z|k) \end{pmatrix}; \quad S_H = \begin{pmatrix} 1 & 0 \\ 0 & 1 \\ 0 & 1 \\ 0 & 0 \end{pmatrix}; \quad R_s = \begin{pmatrix} r_p & 0 \\ 0 & r_l \\ 0 & r_l \\ 0 & 0 \end{pmatrix}.$$

And

$$W = \begin{matrix} \\ p \\ l \end{matrix}\begin{pmatrix} i & j & q & g \\ 0 & w_{pj} & 0 & 0 \\ w_{li} & 0 & 0 & w_{lg} \end{pmatrix} \Rightarrow W^T = \begin{pmatrix} 0 & w_{li} \\ w_{pj} & 0 \\ 0 & 0 \\ 0 & w_{lg} \end{pmatrix}; \ |d(k)> = \begin{pmatrix} d_p(X|k) \\ d_l(Y|k) \end{pmatrix}.$$

Where

$$d_p(X|k) = \Theta(\Delta_{ip}(X|k))\Theta(\delta_{ip}(X|k)),$$

$$d_l(Y|k) = d_{jl}(Y|k)\, d_{ql}(Y|k) = \Theta(\Delta_{jl}(Y|k))\Theta(\delta_{jl}(Y|k))\Theta(\Delta_{ql}(Y|k))\Theta(\delta_{ql}(Y|k)).$$

Then, in the matrix representation:

$$\begin{pmatrix} m_i(X|k+1) \\ m_j(Y|k+1) \\ m_q(Y|k+1) \\ m_g(Z|k+1) \end{pmatrix} = \begin{pmatrix} m_i(X|k) \\ m_j(Y|k) \\ m_q(Y|k) \\ m_g(Z|k) \end{pmatrix} + \begin{pmatrix} -r_p & w_{li} \\ w_{pj} & -r_l \\ 0 & -r_l \\ 0 & w_{lg} \end{pmatrix} \cdot \begin{pmatrix} d_p(X|k) \\ d_l(Y|k) \end{pmatrix}$$

This gives as a result for each mark-function:

$$m_i(X|k+1) = m_i(X|k) - r_p d_p(X|k) + w_{li}(Y, X)\, d_l(Y|k);$$

$$m_j(Y|k+1) = m_j(Y|k) - r_l d_l(Y|k) + w_{pj}(X, Y)\, d_p(X|k);$$

$$m_q(Y|k+1) = m_q(Y|k) - r_l d_l(Y|k);$$

$$m_g(Z|k+1) = m_g(Z|k) + w_{lg}(Y, Z)\, d_l(Y|k).$$

### 5. CHTW-System Classification

At this stage of development of the CHTW-systems concept, it seems appropriate to classify them according to the following properties.

1. Homogeneity

*Homogeneous* – CHTW-systems that have one feature of each classification characteristic, including the same dimension of all branes.

*Heterogeneous* – CHTW-systems that have more than one feature of at least one of the classification characteristics.

2. Topology of CHTW-system

*Linear*, including branching, tree-like**.**

*Network*, including without an explicit direction.

*Radial* (centrifugal and centripetal) – circular, spherical.

*Amorphous* (arbitrary).

*Special* – having a certain functional structure.

3. Stationarity, parametric variability

*Stationary* – the values of CHTW-system parameters are constant over time.

*Non-stationary* - some (or all) values of CHTW-system parameters change when the system is functioning.

*Functionally independent* – CHTW-systems whose parameters don't change or change according to a predetermined law.

*Functionally dependent* – CHTW-systems whose parameters (for example, threshold functions) depend on the state of the system (for example, on the resource distribution in the $i$-th brane).

4. Structural variability

CHTW-systems with a *constant structure*.

CHTW-systems with a *changing structure*.

5. Consideration of uncertainty

*Deterministic* – all variables and parameters of CHTW-system are deterministic.

*Stochastic* - some of the variables and/or parameters of CHTW-system have probabilistic nature.

*Fuzzy* - some of the variables and/or parameters of CHTW-system have a fuzzy representation.

6. Temporality

*Discrete* - the CHTW-system functioning is carried out in steps.

*Continuous* – CHTW-system operates in continuous time.

*Continuous-discrete* (mixed) - part of CHTW-system operates in continuous time, the rest in discrete time.

## 6. Conclusion

The expansion of the standard theory of Petri nets in the direction of accepting the resource multidimensionality inevitably lead to the necessity of both introduction of new concepts and modification of previously accepted ones. In particular, Petri nets transitions and arcs went through a significant reconsideration. W-carriers (the extension of "transition-position" arcs) now acquire a functional sense, transforming the allowed areas of T-branes (transitions) into distributed resources of C-branes, in general, on different spaces. It should also be noted that this work does not implement the transition delay function typical for functional Petri nets [10].

It is impossible to consider many important aspects of the concept of CHTW-systems within the framework of an introductory article. In particular, continuous CHTW- systems, in which time is continuous, remain unconsidered. An important problem awaiting solution: in which numerical systems (**N**, **Z**, **Q**, **R**, **C**) and their combinations is it possible to implement CHTW-systems? For CHTW-systems with a changing structure, the fundamental question is to find the conditions for the fusion and defusion of branes and carriers [4]. Of particular interest is the radial topology of CHTW-systems, which is impossible for traditional Petri nets.

The proposed concept is intended to help researchers and developers carry out more adequate modeling of complex multidimensional systems and technological processes.